\documentclass[manuscript]{aastex}
\usepackage{natbib}
\def\EQN{Eqn. }
\def\158{$\lambda 158\mu$}
\def\63{$\lambda 63\mu$}

\def\nH2{{\rm n}({\rm H}_2)}
\def\NH2{{\rm N}({\rm H}_2)}
\def\pccc{{\rm cm}^{-3}} 

\def\Tstar#1 {\mbox{${\rm T}_{\rm #1}$}}
\def\Tsub#1 {\mbox{${\rm T}_{\rm #1}$}}
\def\TK  {\Tsub K }
\def\TB  {\Tsub B }
\def\DT  {\mbox{$\Delta{\rm T}$}}
\def\TC  {\Tstar C }

\def\TA  {\Tstar A }
\def\Tsys{\Tstar sys }
\def\TheT{\Tstar T }

\def\arcmin{\mbox{$^{\prime}$}}

\def\degr{$^{\rm o}$}
\def\p{$^+$}

\def\h13cop{\mbox{{H$^{13}$CO\p}}}

\def\c3h2{\mbox{C$_3$H$_2$}}
 
 \def\R0{R$_0$}

\def\ddeg{{}^\circ\kern-.1em}

\def\kms{\mbox{km\,s$^{-1}$}}

\def\E#1{\,10^{#1}}
\def\P#1,{$\nH2\TK~=~#1\times~10^4~\pccc$~K}
\def\ec#1,#2,#3,{#1\,(#2)\E{#3}}

\def\H3{\mbox{H$_3$}}

\shorttitle{Noise in the strong-signal limit}

\begin{document}

\title{Noise and the strong signal limit in radio astronomical measurement}

\author{H. Liszt}
\affil{NRAO, 520 Edgemont Rd., Charlottesville, VA 22903-2475}
\email{hliszt@nrao.edu}
%
%
%

%
\begin{abstract}
The random error of radioastronomical measurements is usually computed in 
the weak-signal limit, which assumes that the system temperature is 
sensibly the same on and off source, or with and without a spectral 
line.  This assumption is often very poor.  We give examples of common
situations in which it is important to distinguish the system noise
in signal-bearing and signal-free regions.
\end{abstract}


\section{Introduction}

Few experiments are performed without some attempt at estimating
their errors, and the random errors of measurement in radio
astronomy are typically determined in one general way.  Some form
of comparison is performed whereby samples are taken toward and 
away from a signal source, or with and without a spectral line.  
Subsequent analysis 
proceeds under the assumption that random errors everywhere in the
dataset are as given by the statistical properties manifested in the 
signal-free regions.  No attempt is made to measure the variances of 
signal-bearing and signal-free samples separately during the experiment, 
and, after the fact, random errors of measurement in signal-bearing 
samples are obscured because the form of the signal is arbitrary.  
Discussions of fitting and profile analysis invariably assume that 
measurement variances are the same with or without the signal, as 
for instance the Zeeman analysis of \cite{Mar95} or the fitting of 
functions ($e.g.$ Gaussians) by \cite{KapSmi+66} or \cite{Rie69}. 
Textbook discussions contain no suggestion that system noise is 
influenced by the presence of a signal or that samples with 
different variances may be interleaved in the same datastream
 \citep{Kra86,BurGra01,RohWil00}.

Yet, such treatment has been flawed for a surprisingly long time.  
100 K H I lines have been routinely observed with sub-100 K receiving 
systems for more than 30 years.  Continuum sources whose antenna temperatures
exceed the equivalent noise temperature of the receiving equipment
have been observed even longer.  The error of measurement in signal-bearing 
samples is often significantly different -- with current
receivers it could easily be a factor of 5 at the peak of a 
strong galactic H I emission line -- but the difference has been ignored.
     
Error estimates determine confidence levels and even data containing 
strong signals can be compromised by misunderstanding of their significance; 
for instance,
when two very strong signals are differenced to detect a smaller one
in H I emission-absorption experiments and searches for Zeeman splitting. 
Considering how slowly experimental errors typically improve with the 
amount of time invested in an experiment, it follows that changes 
in the acknowledged errors of an experiment are equivalent to much 
larger differences in the observing time required to reach them.  
$A~priori$ knowledge of errors is an important 
element in the design of experiments and these considerations may 
have a significant effect on the planning of an observing session.  
They should be implemented in the software which supports analysis.

The purpose of this work is to illustrate a variety of common situations 
where random error is dominated by the presence of a
signal.  In the following section some basics of radio astronomy 
measurement are sketched. These are used to analyse 
the statistics of noise and the errors of component fitting when 
signals are present in emission and absorption.  The final
section is a brief summary with an even briefer mention of the
extension of these notions to aperture synthesis.

\section{Power, temperature and noise}

\subsection{Basics}

A temperature scale is established whereby power is compared to the 
classical power spectral density kT (W Hz$^{-1}$) in a resistor in 
thermal equilibrium at temperature T \citep{Dic46,Kra86,RohWil00}.  
The output power level of the telescope system is then quoted as
a \lq system temperature\rq , $i.e.$, k\Tsys.  The actual power 
density  $ h\nu/(e^{h\nu/kT}-1)+h\nu/2$ \citep{CalWel51} 
reduces to kT only in the Rayleigh-Jeans limit and when zero-point 
fluctuations are ignored.

In our simplified discussion we assert \Tsys\ = \TheT + \TA.  
\TheT\ represents everything which does not depend on any particular 
source or input signal and we assume that it is a constant or 
constant function of frequency $\nu$: possible dependencies of 
\TheT\ are suppressed for convenience of notation.  Observing at a
frequency $\nu$ entails a minimum contribution of $h\nu/k$ to \Tsys, 
which is included in $\TheT$.

\TA\ represents a signal external to the telescope.  The  equivalent 
temperature of a signal is its \lq antenna temperature\rq\ which by 
convention is related to the incident flux density 
S$_\nu$ (W m$^{-2}$ Hz$^{-1}$) as S$_\nu$$ = 2$ k\TA/A. The effective 
area A is proportional to the geometric area of the telescope aperture.

The signal may be confined in space or frequency, so we write 
\TA\ = \TA(v) where v is some combination of independent variables.
In the presence of signal the power density is 
k\Tsys(v) =  k\TheT\ + k\TA(v) and the dependence of \TA\ upon v 
makes \Tsys\ 
similarly dependent.  If v represents the pointing of the telescope,
added power comes and goes as the telescope moves. Alternatively, v 
may be velocity or frequency, and, as far as the receiver and square-law 
detector are concerned, the presence of  signal at some v=v\arcmin\ is 
not manifested at v $<>$ v\arcmin.  The passband may be translated or 
inverted by mixing, but the receiver 
and detector electronics are entirely linear in frequency.  The 
spectrum is not jumbled nor is it appreciably smoothed until it is 
integrated and channelized in the so-called \lq backend \rq 
\footnote{Even so, independence of adjacent 1 kHz slices of the 
spectrum, corresponding to 0.2 \kms\ at the H I line, requires a 
minimum integration time of order only 1 msec}.   In Sect. 
2.7 we discuss an exception to this linearity, namely, quantization 
noise in digital correlator spectrometers.

\subsection{Passband or system noise as a measurement variance}

Eventually a datastream is formed from samples of \Tsys, each of duration 
t (say) taken over a spectral width $\Delta\nu$; this could be a spectrum, a 
continuum drift scan, {\it etc.}  Associated with measurement of the output 
power k\Tsys\ there is a variance given by the \cite{Dic46} or 
radiometer equation: 

$$\Delta{\rm T(v)}^2 = {{\Tsys^2({\rm v})}\over{{\rm N}}} = 
 {({\TheT + \TA({\rm v}))^2}\over{{\rm N}}} \eqno(1) $$ 

The dimensionless quantity N $\equiv \Delta\nu {\rm t}$ 
is the product of the bandwidth measured in Hz 
and the integration time in seconds.  Precise determination of 
the output power density kT within a band $\Delta \nu$ is done by averaging 
N independent samples.  Within a band of width $\Delta \nu$ about some 
frequency $\nu$, the contained frequency components beat each other down 
to a frequency range 0..$\Delta \nu$ so that all appear together 
summed within one channel of this width.

Radiometer noise in the output datastream is the measurement variance 
of the power, independent of whether that power was contributed by 
\TheT\ or \TA.  So the variance of the measured strength of an emission 
line, usually considered to be set only by \TheT, actually increases in 
proportion to the source strength itself, weakly for weak signals and more
strongly for very strong ones.

\subsection{Normalization and noise in real-world experiments}

As examples of the way that random error is affected by considerations 
of experimental design, we compare some common methods of data-taking.  
We consider that it is possible to take data ``on''- or ``off''-source; 
if the data are spectra, even the on-source data may have regions of 
the bandpass which are signal-free.

In the simplest case where data are taken while staring at
the source, the variance is given directly by \EQN 1

$$\Delta{\rm T(v)}^2 ~(on)~ = {({\TheT + \TA({\rm v}))^2}\over{{\rm N_{on}}}} .
\eqno(2a)$$ 

When on- and off-source data are differenced the rms is 

$$\Delta{\rm T(v)}^2 ~(on-off)~ = {({\TheT + \TA({\rm v}))^2 }\over{{\rm N_{on}}}} 
 +{{\TheT^2}\over{{\rm N_{off}}}} 
\eqno(2b)$$ 

and the rms in signal-free regions is increased relative to that at
the signal peak.

In some cases a quotient is formed from on- and off-source 
data: the mean off-source power level is equated to a number, 
$\TheT$, and data appear in the form $\TheT$(on/off) or 
$\TheT$(on-off)/off.  Both have the variance

$$\Delta{\rm T(v)}^2 ~(on/off)~ =  {({\TheT + \TA({\rm v}))^2}\over{{\rm N_{on}}}} 
+ {({\TheT + \TA({\rm v}))^2}\over{{\rm N_{off}}}} 
\eqno(2c) $$ 

so formation of the quotient increases the rms in the 
signal-bearing regions relative to the case where simple 
differencing is done, and everwhere relative to the pure ``on'' 
spectrum.

Because of such considerations, it is not possible to calculate 
the random error in signal-bearing regions, even given the empirical rms
in the signal-free regions and the system properties which pertain to
them, unless it is also understood how the data were taken.

%
%
%
%
%
%
%
%
%
%

\subsection{Emission line profiles}

One obvious example of the strong signal limit of a spectral line is
galactic atomic hydrogen.  Fig. 1 shows a typical low latitude galactic 
H I profile observed with a 25m telescope \citep{HarBur97} during the 
Leiden-Dwingeloo Sky Survey (LDSS).  In the lower panel, the scale is 
expanded to show how the $\pm 1 \sigma$ noise envelope varies for 
data taken in the form (on-off)/off with $\TheT = 36$K, a typical value 
during the survey.  The spectrum in Fig. 1 still has very 
high peak/rms signal-noise (465:1), but not nearly as good (1700:1)
as implied by the 0.06 K rms over the baseline regions:
the rms error of the integrated brightness is nearly twice
as high as that estimated from the baseline rms level. 
H I is now commonly observed with $\TheT = 15 - 25 $K.  If the 
same  profile were reobserved with \TheT\ = 18 K for one-fourth 
the amount of time (to reach the same baseline rms ), the 
line-generated rms error would be twice as high again.

From the LDSS, 
we find that some 41\% of the sky contains H I with a peak brightness 
$\TB \ge 20$K, 33\% has $\TB \ge 30$K and 27\% has $\TB \ge 40$K 
(for 0\degr $\le ~l~ \le $ 180\degr, 0\degr $\le ~b~ \le $ 90\degr).

\subsection{Profile fitting}

Discussions of profile fitting typically assume that the rms 
fluctuation is the same in every channel of a spectrum; to do otherwise
would introduce imponderables and greatly hinder general understanding.
However, datapoints having a higher rms should be accorded lower 
weight.

We did a numerical experiment, fitting pure Gaussian profiles
of fixed width (FWHM=$\Delta$V)and varying strength \TA(0), in the
presence of noise which varies following \EQN 1 (a pure ``on'' 
scan following the discussion of Sect. 2.3).  We constructed 
spectra with 1 \kms\ channels at an assumed observing frequency of 
1420.40575 MHz (the $\lambda$21cm line), using \TheT = 20 K typical 
of modern H I receivers.  We assumed an observing time of 30 seconds, 
so that $\sqrt{N} = 376.9$ in \EQN  1 or $\Delta$T = 0.053 K when 
\TA\ = 0.  We then inserted gaussian lines having $\Delta V =
10$ \kms\ and peak strengths \TA(0) = 2.5, 5.0, 10, 20, ... 160 K,
with the variance of the noise in accord with \EQN 1.  Ensembles
of such spectra were generated for each value of \TA(0) and 
fit to single Gaussians. The fitting was done twice for each spectrum, 
weighting by constant or (correctly) changing variance.

The results of this experiment are reported in Fig. 2.  The bottom
curve in each panel is the rms of the fitted parameter given by 
analytic formulae, which coincides with the mean {\it a posteriori} 
error estimate returned by fitting software which assumes a constant 
profile rms.  Stronger lines lead to linear 
improvements in fitting of the central velocity and width in this case, 
while the peak and profile integral fits are independent of strength; 
the fractional precision increases but fitting to the profile integral 
does not achieve higher precision than simply summing the channel values.  

The uppermost curve in each panel is the actual rms of the parameter 
determination with weighting by a constant variance.  The 
shaded (middle) curve is the rms with proper weighting;
in this case, the fitting software returns accurate error estimates.  
Several phenomena are discernible in this diagram.  There are irremediable 
increases in the variances of the fitted parameters relative to the case of 
constant profile rms.  The precision of the fitted centroid and 
width improve only very slowly for strong signals, instead of linearly. 
Variances of the peak and integrated strengths increase in
absolute terms as well.  The fitting is only very slightly improved by 
correct weighting and the actual variances and the claimed error 
estimates diverge sharply if the behaviour of the noise 
is ignored.  This could be misconstrued as implying that the stronger 
lines are less purely Gaussian.

\subsection{Sensitivity of absorption measurements}

Staring at a continuum source characterized by an antenna temperature 
\TA\ = \TC\ results in a system temperature \Tsys = \TheT + \TC.  If 
the continuum is extinguished by a pure scatterer characterized by 
optical depth $\tau(v)$, it follows that

$$\Tsys({\rm v}) = \TheT  + \TC e^{-\tau(v)} \eqno(4) $$

$$\DT({\rm v}) = {{\Tsys({\rm v})}\over{\sqrt{{\rm N_{on}}}}} = 
 {{\TheT +\TC e^{-\tau(v)}}\over{\sqrt{{\rm N_{on}}}}} \eqno(5) $$

The system temperature is higher where there is no absorption.
\EQN 4 can be inverted to solve for the optical 
depth from the observed profile of \Tsys(v), $i.e.$ 
$\tau(v) = -\ln{((\Tsys(v)- \TheT )/\TC)}$.  Neglecting other effects,
the rms of the line/continuum ratio (the argument of the logarithm 
in this expression) is just $\sigma_{l/c} = \DT/\TC$.
$\sigma_{l/c}$ may be normally distributed but 
the logarithmic dependence of $\tau(v)$ makes its error
distribution noticeably asymmetric for moderate to large optical 
depth.  Change in the derived optical depth for a given
fluctuation  $\delta_{l/c}$ in the line/continuum ratio can be written

$$\delta\tau(v)_\mp = \tau(v) + \ln{ (e^{-\tau(v)} \pm \delta_{l/c}) } \eqno(6) $$

where $\pm$ and $\mp$ convey the sense of the variations.
Differentiation yields the rms of the derived optical depth

$$ \Delta\tau(v) = \sigma_{l/c} e^{\tau(v)} 
 [{ {\TheT+\TC e^{-\tau(v)}} \over {\TheT+\TC} }]. \eqno(7)  $$


The usual analysis sets $\tau(v) = 0$ on the right-hand side of \EQN 5
so that the term in brackets in \EQN 7 is unity. 
In Fig. 3 we plot $\delta\tau_+/\tau~ vs.~ \tau$ for different \TC,
taking  $\delta_{l/c} = \sigma_{l/c}$ in \EQN 6 and assuming \TheT = 20 K, 
$\sqrt{N} = 376.9$ as before.  In the usual analysis (upper panel) the 
fractional error in optical depth is minimized at $\tau \approx 1$ for
all \TC\ and sensitivity appears to saturate at rather small 
$\TC \approx \TheT$.  However, use of \EQN 7 shows that the sensitivity 
never saturates, in the sense that it is possible to achieve higher 
precision on ever-thicker lines (lower panel).  Furthermore, the error 
in optical depth at $\tau \approx 1$ is much smaller in the lower panel 
when \TC $>$ \TheT.    


Numerical experiments doing Gaussian fitting to absorption lines
showed (as before) that proper weighting gives slightly improved 
parameter variances, and much-improved error estimates.  Because
the rms is higher in signal-free regions, naive error estimates 
returned by unwitting software are too large.  Error in determining
the continuum level of the baseline regions of an absorption spectrum 
increases with \TC\ and may eventually become the limiting factor in 
determining the line/continuum ratio.


\subsection{Quantization noise}

Use of digital correlators represents a possible departure from the 
frequency-preserving character of the receiver and detector front-end, 
owing to the phenomenon of quantization noise \citep{Ben48,GwiCar+00}.  
Input to the 
correlator is bandpass filtered so that the sampling theorem may be 
applied to recovery of the data,  but digitization of the 
continuously varying input power results in a representation of the
signal which is very strongly {\it not} band-limited.  That portion of 
the power spectrum lying outside the original band is returned, in 
varying degree depending on the sampling rate,
as a form of noise.  For Nyquist sampling (sampling at 
a rate twice the bandwidth) all is returned. For faster sampling 
the return is reduced as sampling sidebands beat with weaker, 
further-out portions of the quantization 
noise spectrum.  As shown by \cite{Ben48} for a 16-level system,
quantization noise is steadily reduced until the sampling rate is 
10 times Nyquist.

Thus, sampling and quantization schemes scatter input power
throughout the passband. Experiments using input thermal noise on
systems with (many) more bits than are used in radio astronomy show 
that the quantization noise is essentially white ({\it ibid}) but 
the spectral characteristics of quantization noise are very much
dependent on the form of the input.  Very strong, highly confined 
signals can produce distortions of the outlying passband.
Weaker signals will simply be dispersed with little
effect on either the noise level or shape of the passband.

Because of quantization noise,  even the blackest absorption line
will not reduce the rms to the level attained in the absence of 
all input signal.  \EQN 5, modified 
to account for quantization loss (1-$\epsilon$) in the case that strong 
absorption occupies a negligibly small fraction of the correlator 
bandpass (so that the quantization noise remains evenly distributed 
over the passband) is

$$ \Delta{\rm T} =  {{\TheT
  + \TC (\epsilon e^{-\tau}+(1-\epsilon))}
\over{\epsilon\sqrt{\rm N}}}. $$

Examples of quantization losses at Nyquist sampling rates are 
(1-$\epsilon$) =  0.36 (1-bit
or 2-level quantization), 0.12 (3-level) and 0.028 (9-level), so that 
minimum fractions $(1-\epsilon)/\epsilon$ = 0.57, 0.14 or 0.03 of the 
rms corresponding to the input \TC\ would unavoidably be present in 
each channel,  including those at the bottom of the line.  This 
complicates error analysis but the high efficiencies of modern 
correlators preserve at least some of the benefits discussed.  
Such considerations are another reason to prefer higher-level 
quantization and over-sampling schemes.

\section{Summary and extension to interferometry}

Radio astronomers frequently observe signals which 
are strong enough to dominate the random errors of their experiments.
Unfortunately, it is not always possible to recognize the effects 
which are induced and they are neglected.  Nonetheless, they have always 
been present in the data.

This discussion points up obvious deficiencies in extant data reduction
software and analysis techniques.  Perhaps less obvious is the need not
only for accurate calibration but also for reliable reporting on the part
of the telescope systems.   Measurement errors cannot be accurately assessed 
and accomodated in downstream data handling unless the system, continuum 
and line antenna temperatures are preserved, along with knowledge of the mode
of data-taking.  Synthesis instruments may be
particularly difficult in this regard.  Consider the use of the VLA
(say) to detect H I absorption against a continuum source at low galactic
latitude in the presence of an emission profile like that shown in Fig. 1.
The VLA does not return the total power or singledish spectra, or,
equivalently, the variation of \Tsys\ across the passband.  The
interferometer experiment {\it per se} can only succeed to the extent
that foreground emission disappears; only its added noise
contribution remains.

We began the discussion by pointing out that the noise contributed 
from sky signals in single-dish observations occurs -- ignoring 
sidelobes, quantization noise and the like -- at those places and/or 
frequencies where the sources themselves are located.  It is an 
interesting endeavour to try to understand the extent to which source 
noise in interferometer experiments is similarly localized in the output 
datastream.  For phased arrays it would seem possible to reproduce the 
single-dish mode.  For synthesis arrays \citep{AnaEke+89,CraNap89} the 
situation is much more complicated
and uncertain even in the weak signal limit.


\acknowledgements

The National Radio Astronomy Observatory is operated by AUI, Inc. under a
cooperative agreement with the US National Science Foundation.  I thank
Darrel Emerson, Tony Kerr, Robert Lucas and A. R. (Dick) Thompson for 
helpful comments.  Barry Clark pointed out the relevance of quantization 
noise and Fred Schwab provided the reference to \cite{Ben48}. This paper was 
put in final form while the author was enjoying the hospitality of the IAP 
in Paris.


\clearpage

\begin{figure}
\plotone{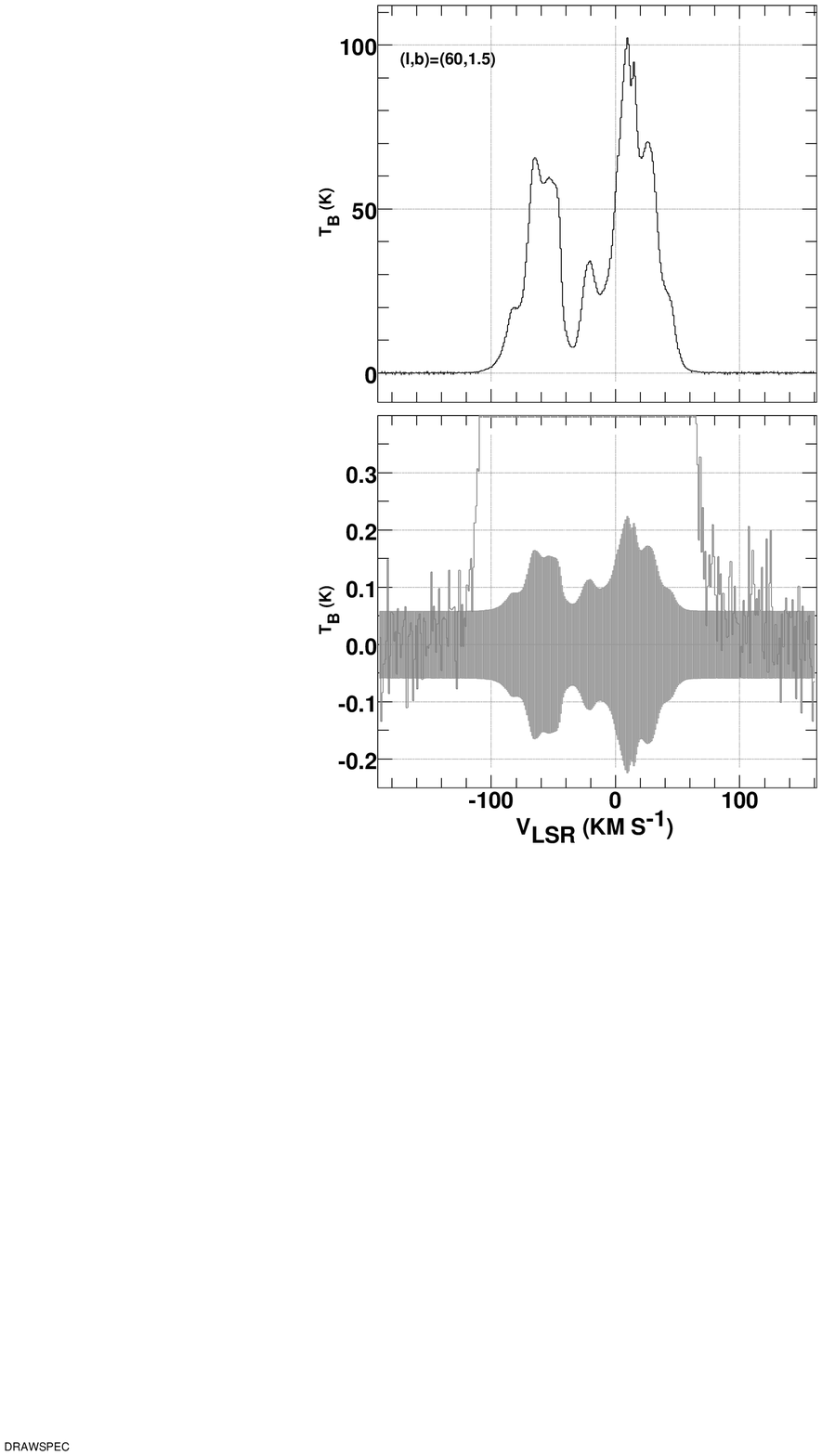}
\caption[]{Top: H I emission observed at (l,b) = (60\degr,$+$1.5\degr)
with the 25m Dwingeloo telescope by \cite{HarBur97}.  Bottom: expanded
view of the $\pm 1 \sigma$ noise envelope assuming $\TheT = 36$ K.}
\end{figure}

\begin{figure}
\plotone{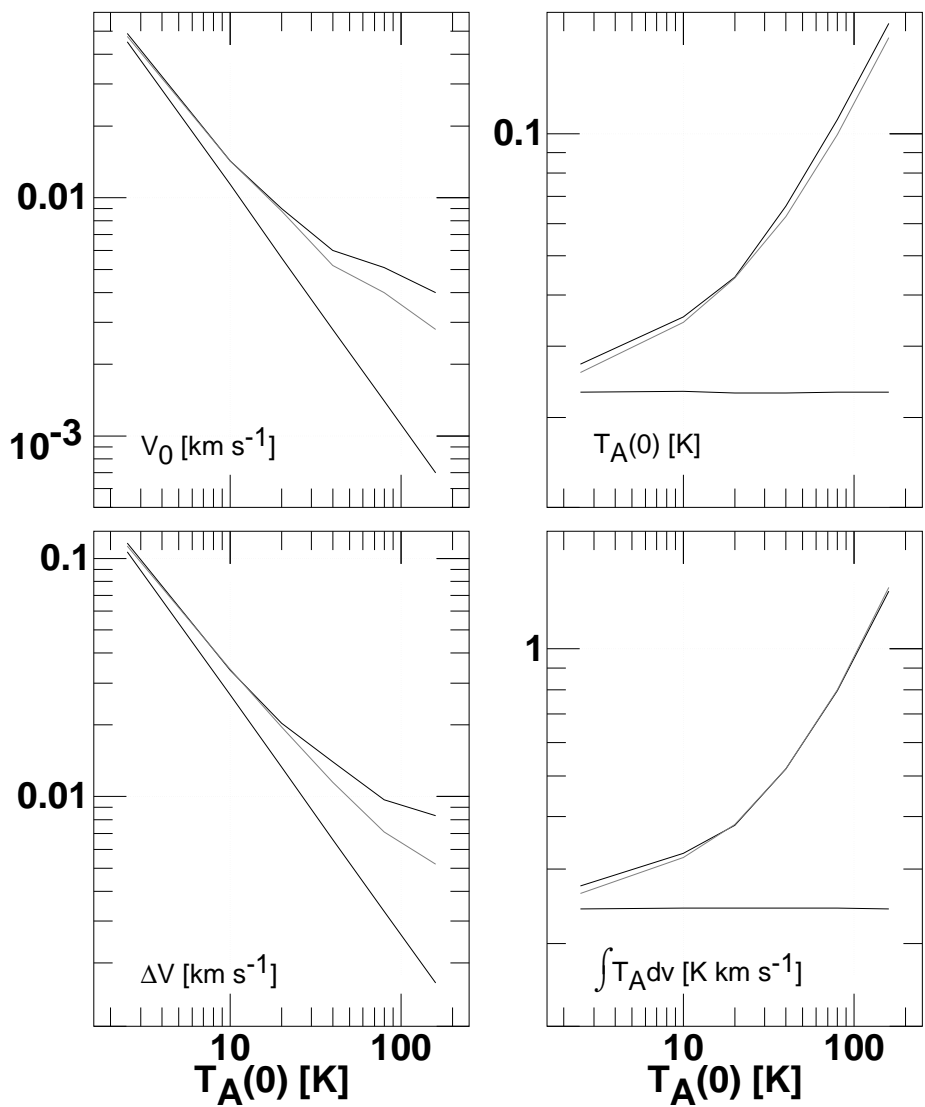}
\caption[]{Rms error of derived gaussian fitting parameters.  Top left,
central velocity.  Bottom left, the FWHM, $\Delta V$.  Top right, the
peak line strength.  Bottom right, the profile integral.  In each panel
the uppermost curve is the empirically-determined rms and the bottommost 
curve the expected or reported rms, all for weighting by a constant
profile variance.  The middle curve is the parameter rms when weighting
by the correct noise variance.  For the other
assumptions used to calculate these curves, see Sect. 2.5}
\end{figure}
\clearpage
\begin{figure}
\plotone{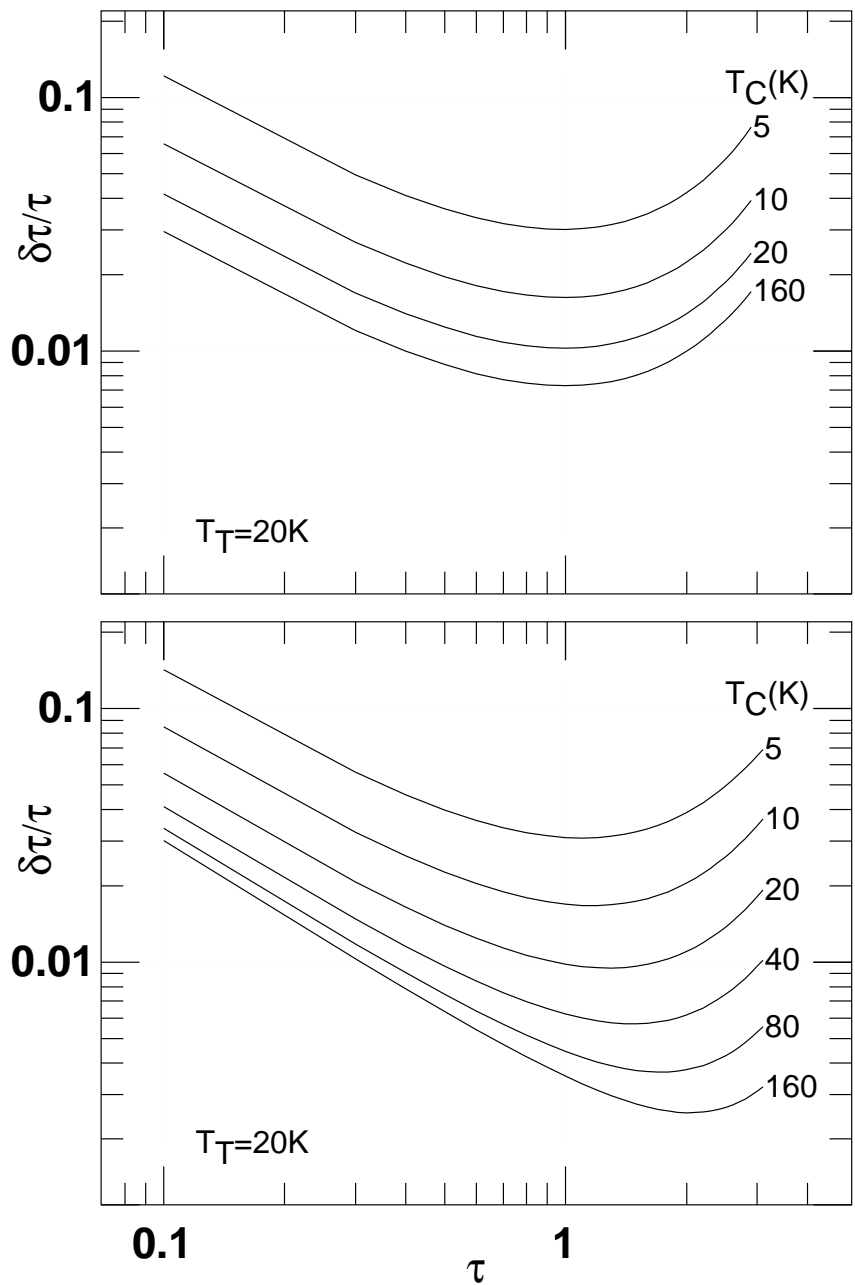}
\caption[]{Fractional rms error in optical depth when a nominal 20 K system 
is used to observe continuum sources of varying strengths 
\TC, occulted by a pure scattering medium of optical depth $\tau$.  
These plots correspond roughly to 30 second integrations
in 1 \kms\ channels at 1420 MHz.  At top, \Tsys\ is assumed independent
of $\tau$; at bottom the dependence of \Tsys\ on $\tau$ (\EQN 7) is 
included.}
\end{figure}
\clearpage

\end{document}